\documentclass[12pt]{article}
\usepackage[dvips]{graphicx}
\usepackage{amsmath}
\usepackage{amsfonts} 
\usepackage{latexsym}
\usepackage{amssymb}
\usepackage{multicol}
\usepackage{array}
\newcommand{\be}{\begin{equation}}
\newcommand{\ee}{\end{equation}}
\newcommand{\bea}{\begin{eqnarray*}}

\newcommand{\eea}{\end{eqnarray*}}
\newcommand{\pa}{\partial}
\renewcommand{\ll}{\left(}
\newcommand{\rr}{\right)}

\newcommand{\gp}{\tilde{G}^+}
\newcommand{\gm}{\tilde{G}^-}

\def\a{\alpha}
\def\b{\beta}
\def\c{\chi}
\def\d{\delta}

\def\g{\gamma}

\def\k{\kappa}
\def\l{\lambda}
\def\m{\mu}

\def\q{\theta}

\def\v{\varphi}

\begin{document}
\thispagestyle{empty} 
\rightline{YITP-SB-04-36}
\vskip 1.5 truecm
\centerline{\LARGE  $N=4$
Superconformal Symmetry} 
\vspace{0.3cm} 
\centerline{\LARGE for the Covariant Quantum
Superstring} 
\vspace{1cm}
\centerline{
\bf{
P.A. Grassi\footnote{pgrassi@insti.physics.sunysb.edu}$^{,a,b}$
and P. van Nieuwenhuizen\footnote{vannieu@insti.physics.sunysb.edu}$^{,a}$
}
}
\vspace{.5cm}
\centerline{$^{(a)}$ 
{\it C.N. Yang Institute for Theoretical Physics,} }  
\centerline{\it State University of New York at Stony Brook,   
NY 11794-3840, USA}  
\medskip
\centerline{$^{(b)}$ {\it Dipartimento di Scienze,
Universit\`a del Piemonte Orientale,}}
\centerline{\it
P.za Ambrosoli 1, Alessandria,  15100, ITALY}

\vskip 2cm
\begin{center}
\begin{minipage}[t]{12cm}
\centerline{\bf Abstract}
\vskip .2cm 
We extend our formulation of the
covariant quantum superstring as a WZNW model with $N=2$
superconformal symmetry to $N=4$. The two anticommuting BRST
charges in the $N=4$ multiplet of charges are the usual BRST
charge $Q_S$ and a charge $Q_V$ proposed by Dijkgraaf, Verlinde and
Verlinde for topological models. Using our recent work 
on "gauging cosets", we then construct a further charge $Q_{C}$ 
which anticommutes with $Q_{S} + Q_{V}$ and which is
intended for the definition of the physical spectrum.  
\end{minipage}
\end{center}
\vskip0.5cm

\eject\vfill

\section{Introduction and Conclusions}

The past four years a new approach to the covariant quantization of the superstring 
has been developed. The starting point is a BRST operator 
$Q_{B} = \oint \l^{\a} d_{z\a}$ in the left-moving sector of the superstring 
\cite{Berkovits:2000fe}, depending on 
free spacetime coordinates $x^{m},\theta^{\alpha}$ and their conjugate momenta 
$p(\theta)_{z\a}$ ($m=0,\dots,9$; $\a=1,\dots,16$),  
and commuting ghosts $\lambda^{\alpha}$. The constraints 
$d_{z\a} \approx 0$ define the conjugate momenta of $\theta^\a$, and this is the only 
information of the classical Green-Schwarz string that is kept. The OPE's 
produce further currents $\Pi_{z m}$ and 
$\partial_{z}\theta^{\alpha}$. Nilpotency of $Q_{B}$ can be achieved by imposing 
the pure spinor constraint $\l \g^m \l =0$, 
but in our approach we have relaxed this constraint, and this produced  
new ghost pairs $(\xi^{m}, \beta_{zm})$ (anticommuting) and 
$(\chi_{\a}, \kappa^{\a}_z)$ (commuting) and a conjugate 
momentum $w_{z \a}$ for $\lambda^\a$ 
(we supress from now on the index $z$ most of the time).
We discovered in this approach that the superstring is a "gauged" WZNW 
model \cite{Grassi:2003kq}, based on a non-semisimple 
nonreductive superalgebra ${\cal A}$ which decomposes into 
coset generators $Q_{\a}$ (associated with $d_{\a}$ and 
$\lambda^{\a}, w_{\a}$) and abelian subgroup generators, namely 
$P_{m}$ (associated with $\Pi_{m}$ and $\xi^{m}, \beta_{m}$) and 
fermionic central charges $K^{\a}$ (associated with $\partial \theta^{\a}$ 
and $\chi_{\a}, \kappa^{\a}$). The matter currents 
$J_{M} = \{d_{\a},\Pi_{m}, \partial \theta^{\a} \}$ depend only on 
$x^{m}, \theta^{\a}$ and $p_{\a}$, and generate ${\cal A}$. 
The gauging leads to a second set of such matter currents $J_{M}^{h}$ 
depending on new variables $x^{m}_{h}, \theta^{\a}_{h}$ and $p^{h}_{\a}$. 
In terms of these currents a particular 
superconformal algebra was constructed, 
with BRST charge $j^{W}_{z B}$ containing the sum 
of the currents $J_{M} + J^{h}_{M}$, a stress tensor, a ghost current, and a spin 2 field 
$B_{zz}$ which contains the difference of the currents $J_{M} - J^{h}_{M}$. 
The central charge of this system vanishes. 

Some preliminary study suggested to us that a second BRST charge was 
needed to define the physical spectrum. We called this unknown charge $Q_{C}$ 
(with C for constraints).\footnote{{A similar analyis has been performed by 
M. Chesterman \cite{Chesterman:2002ey}.}} 
Two BRST charges suggest the presence of an N=4 algebra, but then one 
would expect that the superconformal algebra of the WZNW 
model should be a N=2 subalgebra. We found it not to be  an N=2 algebra, but rather a Kazama 
algebra \cite{Kazama:1990gv}; however, it is known that one can add a gravitational topological quartet 
(which we call the Koszul quartet\footnote{This quartet consists of the ghosts 
$(c^{'z},b^{'}_{zz},\gamma^{'z},\beta^{'}_{zz})$ with conformal spins $(-1,2,-1,2)$ and 
ghost charges $(1,-1,2,-2)$. Later we introduce a second quartet $K= (c^{z}, b_{zz}, \gamma^{z}, \beta_{zz})$ 
with same quantum numbers. } $K'$ below) and modify the Kazama currents 
such that the sums of the currents of the combined system form a genuine $N=2$ 
algebra \cite{figueroa,Getzler:fs}. In particular the BRST charge of this combined 
system is the sum of the separate charges, 
$Q^{W}_S+ Q^{K'}_S$, but the spin 2 current $B^W$ of  WZNW model is modified into 
$\tilde B^W$ by adding terms depending on the fields of $K'$.

The fact that such a Koszul quartet is a gravitational topological quartet was welcome news, 
because it enables us to introduce worldsheet diffeomorphisms into our work. 
It is known from the work of Dijkgraaf-Verlinde-Verlinde \cite{DDV}, that there exist two BRST 
charges in topological models: a charge $Q^{K}_{S}$ for the Koszul quartet 
and a charge $Q_{V}$ which is related to diffeomorphisms and which has the form 
$Q_{V} = \oint c (T^{W} + \frac{1}{2} T^{K}) + \g (\tilde B^W + \frac{1}{2} B^K) + \dots $.  
Here $T^{W}$ is the stress tensor of the matter topological system which in our case 
corresponds to the WZNW system, and $\tilde B^W$ is the modified spin 2 field 
mentioned above. The two charges $Q_{V}$ and $Q^{K}_{S}$ anticommute. 

However, as noticed recently \cite{Guttenberg:2004ht}, 
in order that $Q_{V}$ and $Q^W_S + Q^{K'}_S$ anticommute, 
the Koszul quartet needed to turn the Kazama algebra into N=2 algebra 
cannot be the same as the Koszul quartet  needed to construct $Q_{V}$.  
Thus there are two Koszul quartets, which we already denoted above 
by $K'$ and $K$. The 
quartet $K'$ modifies the current $B^W$ of the WZNW model, while 
$K$ enters in the construction of $Q_{V}$. At this point we have the following 
BRST charges: $Q_{S}^{W}+ Q^{K'}_S$, $Q_{S}^{K}$ and $Q_{V}$ 
where $Q_{V} = 
\oint c (T^{W} + \frac{1}{2} T^{K'}) + \g (\tilde B^W + \frac{1}{2} B^{K'}) + \dots $. 
. The first one is 
a spacetime object, while the latter two are worldsheet objects. They all anticommute. 

Although we have now constructed three BRST charges, none of
them contains the information that the theory originally contained the pure
spinor constraints. So the problem of finding the BRST charge $Q_{C}$ remained. We decided to start a study of general Lie algebras and constraints of the kind  
encountered in the superstring \cite{Grassi:2004cz}. In this study we divided generators into the commuting 
set of Cartan generators, and coset generators. The superstring is an example, 
with $Q_{\a}$ the coset generators, and $(P_{m},K^{\a})$ the abelian subalgebra.  
Since $[Q,P] \sim K$ this algebra is nonreductive as well as 
nonsemisimple. We then ``gauged the coset generators''. By this provocative 
statement we meant that we imposed constraints on the ghosts associated to the 
coset generators (corresponding to the pure spinor constraints \cite{Berkovits:2000fe}
), and then relaxed these constraints in such 
a way that the cohomology remained unchanged. In the process we found 
the second BRST charge $Q_{C}$, but one has to introduce a doubling 
of the subgroup ghosts as well as an another copy of the subgroup ghosts 
which vanishing ghost number. In our case these new fields are denoted by 
$(\xi'_{m}, \beta^{'m}_z, \chi_{\a}', \kappa^{' \a}_z)$ and 
$(\phi_{m}, \bar\phi^{m}_z, \phi_{\a}, \bar\phi^{\a}_z)$. There is a separate BRST 
charge for the coset fields which we denote by $Q_S^{co}$ and a contribution 
of the coset fields to $Q_V$ which we denote by $Q_V^{co}$.  

Following the procedure of \cite{Grassi:2004cz}
the BRST charge $Q^{W}_{C}$ for the WZNW model with $K'$
quartet was recently constructed in \cite{Guttenberg:2004ht}, 
but it was found not to anticommute with the 
total charge ${\bf Q}_S + {\bf Q}_V$ where 
${\bf Q}_S = Q_{S}^{W} + Q^{K'}_{S} + Q^{K}_{S} + Q_S^{co}$ and 
${\bf Q}_V = Q_V + Q_V^{co}$. 
We construct below a charge $Q_{C}$ which does anticommute with ${\bf Q}_{S} + {\bf Q}_{V}$. 
Our construction is based on the observation that all currents so far have been 
constructed without bosonization, so that the zero modes 
$\oint \eta_{z}$ and $\oint \eta'_{z}$ due to bosonization of the 
superghosts of the two Koszul quartets $K$ and $K'$ 
trivially anticommute with all the 
other currents. We propose to take the zero mode
$\oint \eta'_{z}$ and to make a similarity transformation 
with the whole BRST charge ${\bf Q}_{S}$ as follows 
${\bf Q}^W_C = e^{-R} \oint \eta'_{z}  e^{R}$ where 
$R = \{{\bf Q}_{S}, \oint \xi' X^W_{z}\}$. Here $\xi'$ is the partner of $\eta_z'$ and 
$X^W_z$ is defined by $[{\bf Q}_S, \oint X^W_z] = Q^W_C$ with $Q^W_C$ the charge 
given in \cite{Grassi:2004cz}. 
%
%
Of course ${\bf Q}_S$ itself remains unchanged under this similarity transformation 
and ${\bf Q}_C^W$ is of the form $\oint \eta'_{z} + Q_C^W + \dots$ 
and is independent of $K$. The extra 
terms denoted by .... are needed in order that ${\bf Q}^W_C$ 
anticommute with ${\bf Q}_S$. 

Having constructed the extra charge ${\bf Q}_C^{W}$ which we expect 
to be needed to define the correct physical spectrum, we return to the issue of an $N=4$ 
superconformal algebra. A small $N=4$ superconformal algebra needs a triplet of 
$SU(2)$ currents, 
which for a twisted model (the case we are considering) have spins 
$(0,1,2)$ and ghost numbers $(2,0,-2)$ \cite{Berkovits:1994vy}. 

We use the free fields of the $K$ quartet to construct the 
Wakimoto representation of these $SU(2)$ currents \cite{Wakimoto:gf}. 
There are now at least two ways to proceed: 
use ${\bf Q}_{S}$ and ${\bf Q}_V$, or use $Q^{W}_S + Q^{K'}_S + Q^{co}_S$ and ${\bf Q}_C^W$ to 
construct another N=4 algebra. In this paper we perform the first construction. It may clarify 
if we summarize the various charges in a diagram, and indicate the various $N=2$ and $N=4$ 
subalgebras which could conceivably be constructed. Those whose existence 
is only conjectured are indicated by a question mark. From this picture another 
conjecture emerges: the various N=4 algebras are all subalgebras of an enveloping 
$N=8$ superconformal algebra. 

$$\hspace{-2.4cm}\vspace{-.5cm}
\quad{\rm Without \,\, Coset \,\, Fields} \quad \quad \quad 
{\rm With \,\, Coset \,\, Fields}
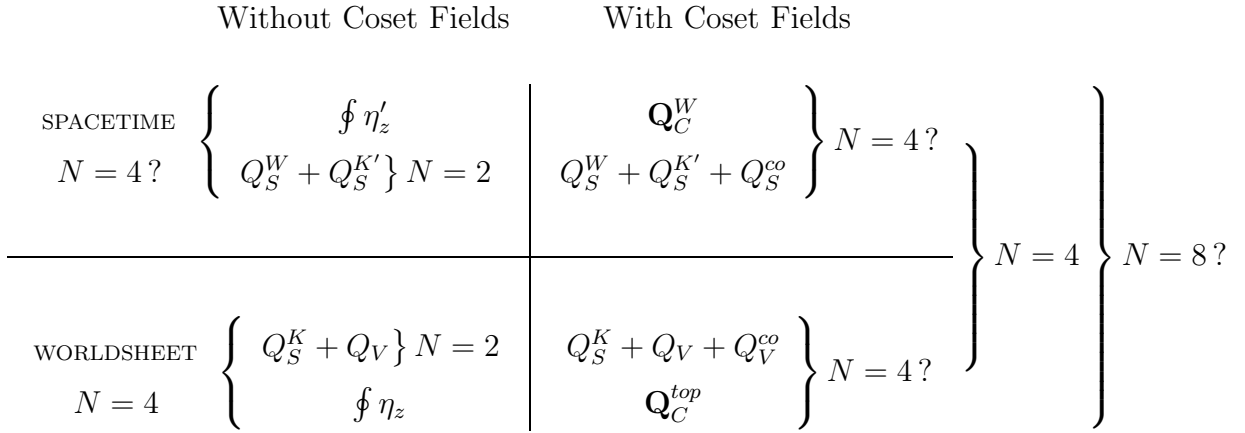
$$
$$$$$$
\vspace{.4cm}
\begin{array}{c|c}
\begin{array}{c} _{\rm  SPACETIME}  \\ N=4\,? \end{array}
\left\{\begin{array}{c}   \oint \eta'_z 
\\ \left.  Q^W_S + Q^{K'}_S \right\} N=2  \end{array}\right. &  \left. \begin{array}{c}
{\bf Q}^W_C \\ Q^W_S + Q^{K'}_S  + Q_S^{co} \end{array} \right\} 
N=4\,?\\\\ \hline \\ \begin{array}{c} _{\rm WORLDSHEET}  \\
N=4\end{array} \left\{\begin{array}{c} \left.  Q^K_S + Q_V \right\} N=2 \\ 
\oint \eta_z \end{array}\right. &   \left. \begin{array}{c} Q^K_S + Q_V  + Q^{co}_V \\ 
{\bf Q}^{top}_C \end{array} \right\} N=4\,? \end{array}  \left. 
\vphantom{\begin{array}{c} a \\ b \\ c \\ d  \end{array}} \right\} 
N=4 \left. \vphantom{\begin{array}{c} a \\ b \\ c \\ d \\ e \\ f \end{array}} 
\right\} N=8\,? 
$$
$$\vspace{.2cm}
{\bf Mutually \,\, Anticommuting\,\, BRST \,\, Charges\,\, of\,\, N=2,4\,\, Subalgebras}
$$

In the spacetime sector we begin with the BRST charge $Q^W_S$ of the WZNW model 
\cite{Grassi:2003kq} (see the left upper part of the diagram). The BRST charge $Q^W_S + Q^{K'}_S$  
belongs to an N=2 algebra \cite{Grassi:2003kq}. The BRST charge\footnote{The bosonization 
formulas for $K$ are $\gamma^z = \eta_z e^{- \varphi}$ and $\beta_{zz} = \partial \xi e^{\varphi}$ with 
$\varphi(z) \varphi(w) \sim - ln(z-w)$. Similarly for $K'$.} $\oint \eta_z'$  anticommutes 
$Q^W_S + Q^{K'}_S$ and these two charges might be part of $N=4$ algebra. The coset fields are 
needed to construct $Q_C^W$ according to 
\cite{Grassi:2004cz} and hence one finds the BRST charge $Q^{co}_S$ for the coset fields in the 
right upper part of the diagram. 
Comparing the left- and right-hand side of the diagram, 
we conjecture that the BRST charges ${\bf Q}^W_C$ 
which we discussed above and $Q^W_S + Q^{K'}_S + Q^{co}_S$ are part of an another 
$N=4$ algebra. 
 
In the worldsheet sector we find the BRST charge $Q_S^K + Q_V$ which 
is part an N=2 algebra, as discussed in \cite{DDV}, see the lower left part of the 
diagram. The zero mode $\oint \eta_z$ forms another anticommuting BRST charge, and 
together these two BRST charges form an N=4 superconformal algebra as shown by Berkovits and 
Vafa \cite{Berkovits:1994vy}. We can repeat our procedure in the 
spacetime sector and make a similarity transformation on $\oint \eta_z$ 
with the BRST charge of the worldsheet sector to obtain ${\bf Q}^{top}_C$, see the lower 
right part of the diagram. 
The formula reads ${\bf Q}^{top}_C = e^{-R^{top}} \oint \eta_{z} e^{R^{top}}$ where 
$R^{top} = \{Q^{K}_{S} + {\bf Q}_V, \oint \xi X_{z}^{top}\}$ and $X_{z}^{top} = c' b$. 
The construction of $X_{z}^{top}$ follows along 
the lines of \cite{Grassi:2004cz}: one starts 
with the Lie derivative along $\gamma$ \cite{top_witt}, which is the analog of the constraints 
involving $d_\a$, and then one relaxes this constraint \cite{Ouvry:1988mm}. In this context $\gamma$ plays the role 
of $\lambda^\a$, $c$ plays the 
role of $(\xi^m, \chi_\a)$, while $c'$ corresponds to $(\xi^{'m}, \chi'_{\a})$, and $\gamma'$ 
corresponds to $(\varphi^m, \varphi_\a)$. Note that "gauging" of the coset of the topological quartet $K$ brings in the quartet  $K'$ of the spacetime sector. This is the more fundamental reason 
that one needs two quartets. Perhaps again $Q^K_S + {\bf Q}_V$ and ${\bf Q}^{top}_C$ are part 
of an N=4 algebra (see the lower right part of the diagram). 

Finally, we come to the contents of this paper. We show that ${\bf Q}_S$ and 
${\bf Q}_V$ do indeed belong to an N=4 superconformal algebra. We construct this algebra in steps. 
In section 2 we construct an N=4 algebra for the quartet $K$ with the Wakimoto triplet, in section 3 we add the 
coset fields, and in section 4 we add the WZNW model coupled to the quartet $K'$. The final result is given 
by eq. (45).  

It is also easy to construct a charge $Q_{C}$ which anticommute with ${\bf Q}_{S} + {\bf Q}_{V}$, 
namely $Q_{C} = e^{-R} \oint (r \eta + s \eta') e^{R}$ with arbitrary $r,s$, and 
$R = \{{\bf Q}_S + {\bf Q }_V, \oint X\}$. One choice for $X$ is $X=\left( \xi' X^W_z + \xi X^{top}_z\right)$. 
In order that physics after the similarity transformation 
is different from physics before, we expect that  a suitable filtration (grading condition 
\cite{grassi}\footnote{See \cite{Grassi:2004cz} for a geometrical 
interpretation of the grading.}) is needed.  

Despite several important results of the pure spinor formalism \cite{psf} 
obtained by N. Berkovits and the Stony Brook group, a deeper
understanding of the formalism and its geometrical origins are 
still lacking. Several issues such as the relation with the 
kappa symmetry of Green-Schwarz string theory, 
the Virasoro constraints (and therefore the diffeomorphism invariance), 
and the role of picture changing operators in a path-integral 
construction have to be explored and the present paper might 
shed some light on these aspects.

  
\section{The $N=4$ gravitational Koszul quartet.}

The quantization of the superstring as a particular WZNW model
based on a nonsemisimple Lie algebra has led us in~\cite{Grassi:2003kq} to a
twisted $N=2$ superconformal field theory. 
Following~\cite{DDV} we
introduced a gravitational $N=2$ Koszul quartet which can be
considered as the twisted version of the familiar spin
$(2,-1,3/2,-1/2)$ ghost quartet $(b_{zz},c^z,\b_{zz},\g^z)$ 
of the $N=1$ RNS spinning string. The introduction of this quartet
achieved two goals:

\begin{itemize}
\item[(i)] it allowed the construction of a second BRST
charge $Q_{V}$ as in topological models, and
\item[(ii)] it coupled our spacetime-supersymmetric superstring to
worldsheet gravity.
\end{itemize}
The two BRST charges $Q^{W}_{S}$ and $Q^{K}_{S}$ 
are present in any topological model, so
they can not be used to eliminate the dependence on $x_h$ and
$\q_h$. We need another anticommuting operator, like an 
antighost, to eliminate this dependence. Moreover, if one has two BRST
charges, it seems likely that one is dealing with an $N=4$ model.

An $N=2$ model with two spin-one BRST charges suggest that it is
part of a twisted $N=4$ model, which should consist of two
spin-one BRST currents $G^+(z)$ and $\tilde{G}^+(z)$, two spin 2
$B$-fields $G^-(z)$ and $\tilde{G}^-(z)$, a stress tensor $T_{zz}$
with vanishing anomaly, and further $SU(2)$ currents. In a twisted
$N=4$ model the $SU(2)$ currents have spin $0,1,2$, rather than
spin 1~\cite{Berkovits:1994vy}. We thus need a spin $(0,1,2)$ triplet of
currents which separately form a closed algebra.

At this point we may recall that the well-known Wakimoto
representation~\cite{Wakimoto:gf} of currents constructed from ghost fields
satisfies these properties. One is thus led to study the original
$N=2$ gravitational Koszul quartet together with the Wakimoto
triplet of currents, and try to extend this model to an $N=4$
model.

This quartet $(b_{zz},c^z,\b_{zz},\g^z)$ has spins
$(2,-1,2,-1)$ and ghost charges $(-1,1,-2,2)$. The ghosts $b_{zz}$
and $c^z$ are anticommuting with propagator
$c(z)b(w)\sim(z-w)^{-1}$, while $\g^z$ and $\b_{zz}$ are commuting
and satisfy the OPE $\g(z)\b(z)\sim(z-w)^{-1}$. The currents of
this $N=2$ model are given by
\begin{align}
T_{zz}&=-2b_{zz}\pa_zc^z-\pa_zb_{zz}c^z-2\b_{zz}\pa_z\g^z-\pa_z\b_{zz}\g^z\\
j_z^B&=-b_{zz}\g^z,\quad J_z =-b_{zz}c^z-2\b_{zz}\g^z\\
B_{zz}&=2\b_{zz}\pa_zc^z+c^z\pa_z\b_{zz}+\m b_{zz}
\end{align}
The stress tensor is simply the sum of the stress tensors of two
spin $(2,-1)$ doublets, and the factor 2 in the ghost current
yields the ghost charges $(2,-2)$ for $\g^z$ and $\b_{zz}$. The
$B$ field $B_{zz}$ has spin 2 and ghost number $-1$, and the
parameter $\m$ is a free parameter (to be fixed to $\mu=1$ later).
The spin-1 BRST current $j_z^B$ and the spin-2 field $B_{zz}$ are
the twists of the two spin $3/2$ currents of an untwisted $N=2$
multiplet. From now on we shall drop the super- and subscripts $z$
when no confusion is possible.

The Wakimoto representation is given by
\begin{align}
J^{++}&=-bc\g+\frac32\pa\g-\b\g\g\\
J_3&=-bc-2\b\g\\
J^{--}&=\b
\end{align}
The superscripts denote the ghost number. 
The ghost current is identified with $J_3$. These currents satisfy
the following OPE
\begin{align}
&J_3(z)J^{\pm\pm}(w) \sim\pm 2\frac{J^{\pm\pm}(w)}{z-w};& &J_3(z)J_3(w)\sim\frac{-3}{(z-w)^2}\\
&J^{++}(z)J^{--}(w) \sim\frac{-3/2}{(z-w)^2}+\frac{J_3(w)}{z-w};&
&T(z)J_3(w)\sim\frac{3}{(z-w)^3}+\frac{J_3(z)}{(z-w)^2}
\end{align}
Closure of the algebra fixes all coefficients in the currents. We
could rescale these currents such that the terms with double poles 
in $J^{++} J^{++}$ and $J_{3} J_{3}$ 
become equal, but the formulas are simpler by keeping the present
normalization.

We now present the $N=4$ extension of the $N=2$ Koszul model. 
This result has been obtained before in \cite{oldN=4} 
with $\mu =0$, but we keep $\mu$ arbitrary. For completeness we give the derivation with 
$\mu\neq0$. The
stress tensor and $SU(2)$ triplet  are unchanged, while we have
the following anticommuting currents
\begin{align}
G^+=-b\g \ \ &\xleftarrow{J^{++}}\xrightarrow{J^{--}}   \tilde{G}^-=-b \\
 G^-=2\b\pa c + c\pa\b+\m b\ \ &\xleftarrow[J^{--}]{}
 \xrightarrow[J^{++}]{}
 \tilde{G}^+ = -\frac32\pa^2c+bc\pa c +2\pa c\b \g +
c\pa\b\g+\m b \g
\end{align}
The currents $G^\pm$ are equal to the BRST current and the $B$ field 
of the $N=2$ model. 
As the notation indicates the currents $J^{++}$ and $J^{--}$ map
the currents $G^+$ and $\gm$ into each other, and also $G^-$ and
$\gp$ are mapped into each other by $J^{++}$ and $J^{--}$
\begin{align}
&J^{++}(z)G^+(w)\sim 0;& &J^{--}(z)G^-(w)\sim 0\\
&J^{++}(z) \gp(w)\sim 0; & &J^{--}(z)\gm(w)\sim 0\\
&J^{++}(z)G^-(w)\sim \frac{-\gp(w)}{z-w};& &J^{--}(z)\gp(w)\sim \frac{-G^-(w)}{z-w}\\
&J^{++}(z)\gm(w)\sim \frac{-G^+(w)}{z-w};& &J^{--}(z)G^+(w)\sim
\frac{-\gm(w)}{z-w}
\end{align}
Only the calculation of $J^{++}(z) \tilde G^{+}(w)$ is involved. 

The superscripts of these currents denote their ghost number \be
J_3(z)G^{\pm}(w)\sim\pm\frac{G^{\pm}(w)}{z-w};\ \ \
J_3(z)\tilde{G}^{\pm}(w)\sim \pm \frac{\tilde{G}^{\pm}(w)}{z-w}
\ee The conformal spin of $G^+$  and $G^{-}$ is 1 and 2,
respectively~\cite{Grassi:2003kq}, 
while it is straightforward to verify
that $\tilde{G}^{\pm}(w)$ have the same conformal spin as
$G^{\pm}$
\begin{align}
&T(z)\gp(w)\sim\frac{\gp(w)}{(z-w)^2}+\frac{\pa\gp(w)}{z-w}\ \ 
\\
&T(z)\gm(w)\sim\frac{2\gm(w)}{z-w}+\frac{\pa\gm(w)}{z-w}
\end{align}

The crucial test is whether the OPE's of two fermionic currents
close. They do indeed close. We find the following OPE's
\begin{align}
&G^+(z)\gp(w)\sim\frac{2J^{++}(w)}{(z-w)^2}+\frac{\pa
J^{++}(w)}{z-w}\\
&G^-(z)\gm(w)\sim\frac{2J^{--}(w)}{(z-w)^2}+\frac{\pa
J^{--}(w)}{z-w}\\
&G^+(z)G^-(w)\sim\frac{-3}{(z-w)^3}+\frac{
J_3(w)}{(z-w)^2}+\frac{T_{zz}(w)}{z-w}\\
&\gp(z)\gm(w)\sim\frac{3}{(z-w)^3}+\frac{
-J_3(w)}{(z-w)^2}+\frac{-T_{zz}(w)}{z-w}
\end{align}
For our work it is important that the two BRST $\oint G^{+}$ and 
$\oint \tilde G^{+}$ charges are
nilpotent and anticommute. This is indeed the case
\begin{align}
&G^+(z)\gm(w)\sim 0; &&G^+(z)G^+(w) \sim 0; &&\gm(z)\gm(w)\sim 0\\
&G^-(z)\gp(w)\sim 0; &&G^-(z)G^-(w) \sim 0; &&\gp(z)\gp(w)\sim 0
\end{align}

For $\tilde G^{+}(z) \gp(w)$ we directly checked that the terms with $\mu$ 
cancel, but the vanishing of this OPE follows already from (13) and (23). 
 
We conclude that we have constructed an $N=4$ extension of the
gravitational $N=2$ Koszul quartet. We end this section with a few
comments
\begin{itemize}
\item[1)] The parameter $\mu$ of the term $\m b$ in $G^-$ remains
arbitrary; it is not fixed when one extends the $N=2$ Koszul model
with a free $\m$ to the $N=4$ Koszul model.
\item[2)]Both $T,J_3,G^+,G^-$ and $T,J_3,\gp,\gm$ are $N=2$
multiplets. Since obviously for both the anomaly in $TJ_3$ is
opposite to the anomaly in $J_3J_3$, both are topological $N=2$
multiplets. The anomaly in the stress tensor indeed vanishes.
\item[3)] The OPE's of a twisted $N=4$ model are for example given
in~\cite{Berkovits:1994vy}. We obtain agrement with these OPE's if we rescale
our current by factors $\pm i$.
\item[4)] For $\mu =0$ this N=4 superconformal algebra has been derived before 
in \cite{oldN=4}, specifically equation (33).
\end{itemize}

\section{An $N=4$ model for one Koszul quartet and coset fields}
In this section we extend the construction to ``coset fields''.
These coset fields were first introduced in our paper~\cite{Grassi:2004cz}, 
in order to construct a second BRST
change for the superstring called $Q_C$. Subsequently these
fields were added to our $N=2$ WZNW model for the superstring
in~\cite{Guttenberg:2004ht}. The result of these articles is an $N=2$
conformal field theory containing two Koszul quartets, coset
fields, and the fields of the WZNW model. In this section we
construct an $N=4$ conformal field theory containing one Koszul
quartet and the coset fields. This will pave the way to an $N=4$
formulation of the WZNW model.

The coset fields for the superstring consist of second set of
ghosts $({\xi'}_m, {\b'}^{m}_{z}, \c'_{\a}, {\k'}^{\a}_z)$, and a
corresponding set of fields $(\v_m,\bar{\v}^m,\v_\a,\bar{\v}^\a)$.
The fields $(\xi'_m,\b'_m,\v_\a,\bar{\v}^\a)$ are  anticommuting,
while $({\c'}_{\a} ,{\k'}_z^\a,\v_m,\bar{\v}^m)$ are commuting.
The propagators are the standard ones
\begin{align}
& \xi'_m(z)\b'^n_{z}(w)\sim \d_m^n\frac{1}{z-w}; &&
{\c'}_\a(z){\k'}_z^\b(w)\sim\d_\a^\b\frac{1}{z-w}\\
&\v_m(z)\bar{\v}_{z}^n (w)\sim\d_m^n\frac{1}{z-w}; && \v_\a
(z)\bar{\v}^\b_{z}(w)\sim\d_\a^\b \frac{1}{z-w}
\end{align}
Following~
\cite{Grassi:2003kq,Grassi:2004cz,Guttenberg:2004ht} the stress tensor, ghost and $B$
field are easily written down. For $T_{zz}$ we have the usual free
field expression
\begin{align}
T^{co+K}=&-{\b'}_{zm}\pa_{z}{\xi'}^m-{\k'}_z^\a\pa_{z}{\c'}_\a-\bar{\v}_{z m}\pa_{z}\v^n-\bar{\v}^\a_{z}\pa_{z}\v_\a \nonumber\\
&-2b_{zz}\pa_{z} c^{z}-\pa_{z} b_{zz} c^{z}-2\b_{zz}\pa_{z}\g^{z}-
\pa_{z}\b_{zz}\g^{z}\ \ \ \mbox{with $c_{TT}=0$}
\end{align}
The central charges of the $bc$ and $\b \g$ system ($-26$ and
$26$) cancel each other, and also those of the coset fields
cancel because the primed fields have opposite statistics from the
$\v$ fields. The ghost current is the sum of the ghost currents of
the two systems \be
J_z^{co+K}=-{\b'}_{zm}{\xi'}^m-{\k'}_z^\a{\c'}_\a-b_{zz}c^z-2\b_{zz}\g^z\
\ \ \mbox{with $c_{JJ}=-9$}\ee Its anomaly is $c_{JJ}=-9$.  
(Twisting yields this anomaly in the $JJ$ \ OPE, while the
conformal anomaly in $TT$ vanishes after twisting). The BRST
current is the sum of the two BRST currents of the coset and
Koszul systems \be
j_{z,B}^{co+K}=-\bar\varphi_z^m{\xi'}_m- \bar\varphi_z^\a{\c'}_\a-b_{zz}\g^z \ee
Finally, the $B_{zz}$ field reads \be
B_{zz}^{co+K}={\b'}_{zm}\pa_{z}\v^m+{\k'}_z^\a\pa_{z}\v_\a+2\b_{zz}\pa_z
c^z+c^z\pa_z\b_{zz}+\m b_{zz} \ee
where we recall that $\mu$ is a free parameter. 

The coset currents $T_{zz}^{co},J_z^{co},j_{z,B}^{co}$ and
$B_{zz}^{co}$ form separately an $N=2$ superconformal algebra. In
particular
\begin{align}
&j_B^{co}(z)B^{co}(w)\sim\frac{-6}{(z-w)^3}+\frac{J^{co}}{(z-w)^2}+\frac{T^{co}}{z-w}\\
&J^{co}(z)J^{co}(w) \sim \frac{-6}{(z-w)^2}\\
&T^{co}(z)J^{co}(w)\sim
\frac{6}{(z-w)^3}+\frac{J^{co}(w)}{(z-w)^2}+\frac{\pa
J^{co}(w)}{z-w}
\end{align}
However, in the extension to an $N=4$ system, couplings arise
between the coset and the Koszul system, as we now show.

To obtain the extension to an $N=4$ system we need to extend the
$U(1)$ ghost current to an $SU(2)$ current triplet with conformal
spin $(0,1,2)$. The following is such a system
\begin{align}
&J^{++}=J_z^{co}\g^z+\frac{9}{2}\pa_z\g^z-\g^z\g^z\b_{zz}-\g^zb_{zz}c^z-c^zj_B^{co}\\
&J_3
=-{\b'}_{zm}{\xi'}^m-{\k'}_z^\a{\c'}_\a-b_{zz}c^z-2\b_{zz}\g^z\\
&J^{--}=\b_{zz}
\end{align}
The ghost values of these currents are $(2,0,-2)$ respectively
\begin{align}
&J^{\pm\pm}(z)J_3(w)\sim\mp2\frac{J^{\pm\pm}(w)}{z-w}\\
&J^{++}(z)J^{--}(w)\sim
\frac{-9/2}{(z-w)^2}+\frac{J_3(w)}{z-w}\\
&J_3(z)J_3(w)\sim\frac{-9}{(z-w)^2}
\end{align}
All coefficients in the $SU(2)$ current are fixed by requiring
closure, in particular the coefficient of 
the total derivative $\frac{9}{ 2} \, \partial_z \gamma^z$.

We can now construct the currents $\gp$ and $\gm$ by acting with
$J^{++}$ and $J^{--}$ on $j_{zB}^{co+K}\equiv G^+_z$ and
$B_{zz}^{co+K}\equiv G_{zz}^-$. One finds easily \be
J^{--}(z)G^+(w)\sim\frac{-\gm(w)}{z-w} \ \Rightarrow \  \gm=-b \ee
The calculation of $\gp$ is more involved. We start from
\begin{align}
&-\frac{\gp(w)}{z-w}\sim J^{++}(z)G^-(w)=\ll J^{co}\g-c
j_B^{co}+\frac{9}{2} \pa\g-\g\g\b-\g bc \rr (z) \nonumber \\
&\ll B^{co}+2\b\pa c+c\pa\b+\m b \rr(w)
\end{align}
We obtain
\begin{align}
\tilde{G}^{+}& =\   cT^{co} +\g B^{co}-\pa\ll cJ^{co}\rr -\m
j_B^{co}
-\frac{9}{2}\pa^2c \nonumber\\
& +bc\pa c +2\g\b\pa c + \g c\pa\b + \mu \gamma b\,.
\end{align}
Triple and double poles nicely cancel here, confirming the
coefficient $9/2$ of the term with $\pa\g$ in $J^{++}$. The crucial
question is whether the simple structure of $\tilde G^{+}$ 
in the coset sector also holds in the Koszul sector. We find
\begin{align}
bc\pa c+2\g\b\pa c +\g c\pa\b=c\ll\frac12T^K \rr+\g\ll \frac12B^K
\rr -\pa \ll c\frac12J^K\rr + {\m  \over 2} j_B^K
\end{align}

Hence, the total $\gp$ is indeed of a simple form
\begin{align}
\gp &= c\ll T^{co} +\frac12 T^K\rr + \g \ll B^{co}+\frac12 B^K \rr
-\pa \ll c\ll J^{co}+\frac12 J^K\rr \rr \nonumber \\
&- \m (j_B^{co}+ \frac12 j_{B}^{K})-\frac92\pa^2c
\end{align}
Also $J^{++}$ can be written in this way \be J^{++} = \g \ll
J^{co} + \frac12 J^K\rr -c \ll j_B^{co}+\frac12 j_{B}^{K}\rr
+\frac92\pa\g \,. \ee 

\section{The WZWN model coupled to two Koszul quartets and coset fields}
In the previous section we saw how an $N=2$ ``matter" system (the
coset fields) could be coupled to a Koszul quartet such that an
$N=4$ model resulted. We only needed the OPE's of the currents of
the matter system. This reveals how to couple the WZWN model to
these fields such that it becomes part of an $N=4$ model
\begin{itemize}
\item[(i)] use a first Koszul quartet denoted by $(b',c',\b',\g')$
to construct a bona fide $N=2$ system for the WZWN model with
currents $T^W,J^W,j^W,\tilde{B}^W$~\cite{Grassi:2003kq}. This fixes the $\m$
parameter of the first quartet to $\m=1$.
\item[(ii)] couple this $N=2$ system to a second Koszul quartet,
denoted by $(b,c,\b,\g)$, to obtain an $N=4$ model in the same way
as for the coset fields. The $\m$ parameter of this Koszul quartet
is arbitrary. Instead of coupling only to the second Koszul quartet 
we shall couple to the sum of the second Koszul multiplet and the 
coset fields. This combined system was discussed in the previous 
section and is what is needed below. 

\end{itemize}
Thus we obtain the following $N=4$ superconformal currents for the
WZWN model coupled to coset fields and two Koszul quartets
\begin{align}
T&=(T^W+T^{K'})+T^{co}+T^K\ \mbox{with $c_{TT}=0$} \nonumber\\
J_3&=(J^W+J^{K'})+J^{co}+J^K\ \mbox{with $c_{JJ}=-22-3-6-3 = -34$}  \nonumber\\
G^+&=j_B=(j_B^W+j_B^{K'})+j_B^{co}+j_B^{K}  \nonumber\\
G^-&=B=(\tilde{B}^{W}+B^{K'})+B^{co}+B^{K}  \nonumber\\
J^{++}&=\g(J^W+J^{K'}+J^{co}+\frac12 J^K)-c(j_B^W+j_{B}^{K'}+j_B^{co}+\frac12j_B^K)+x\pa\g 
 \nonumber\\
\gp&=c (T^W+T^{K'}+T^{co}+\frac12
T^K)+\g(\tilde B^W+B^{K'}+B^{co}+\frac12 B^K) 
 \nonumber\\
&-\m (j_{B}^{W} + j_{B}^{K'} +j_{B}^{co} + \frac12 j_{B}^{K}) -\pa(c(J^W+J^{K'}+J^{co}+\frac12J^K))+y\pa^2c \nonumber\\
J^{--}&=\b;\quad\gm=-b
\end{align}

The current $J^{++,K}$ contains a term $x\pa\g$ while the current
$\gp$ contains a term $y\pa^2c$. The same analysis as performed
for the coset fields shows that also these currents satisfy an
$N=4$ superconformal algebra. The only parameters to be fixed are
the values of $x$ and $y$. We fix $x$ by requiring that the double
poles with $\g$ in the numerator cancel in the following OPE \be
J^{++}(z)J_{3}(w)\sim -2\frac{J^{++}(w)}{z-w}+{\cal
O}\frac{1}{(z-w)^2} \ee We find
\begin{align}
&\left[ 
\g(J^W+J^{K'}+J^{co}+\frac12J^K)-c(j_B^W+j_B^{K'}+j_B^{co}+\frac12j_B^K)+x\pa\g
\right](z) \nonumber\\
&\left[J^W+J^{K'}+J^{co}+J^K
\right](w)\sim\frac{2x\g(w)}{(z-w)^2}+\g(z)\frac{[-22-3-6-(\frac12+4-\frac12)]}{(z-w)^2}+\dots
\end{align}
This yields the value \be x=17 \ee Confirmation is obtained from
\be J_3(z)J_3(w)\sim\frac{-34}{(z-w)^2}; \ \
J^{++}(z)J^{--}(w)\sim\frac{-x}{(z-w)^2}+\frac{J_3(w)}{z-w} \ee
which reproduces $x=17$.

Finally we complete the construction of the $N=4$ WZNW model by
determining the value of $y$. We consider the OPE
$J_3(z)\gp(w)\sim\gp(w)/z-w$ and require that all terms of the
form $c(w)/(z-w)^3$ cancel. We find the following contributions
\begin{align}
&(-bc)(z)(bc\pa c+y\pa^2c)(w)-(2\b\g)(z)(2\b\g\pa c+\pa\b\g
c)(w)\nonumber\\
&+(J^W+J^{K'}+J^{co})(z)(c(T^W+T^{K'}+T^{co}))(w)\nonumber\\
&-(J^W+J^{K'}+J^{co})(z)\pa(c(J^W+J^{K'}+J^{co}))(w)\nonumber\\
&\sim [1+2y+2+(-22-3-6)-2(-22-3-6)]c(w)/(z-w)^3
\end{align}
Thus \be y=-17 \ee As a check we determine the term with $\pa^2 c$
in $\gp$ from $J^{++}(z)G^-(w)\sim -\gp(w)/(z-w)$. We find
\begin{align}
&\left[ \begin{array}{l} \gamma(J^W+J^{K'}+J^{co})+\frac12\g(-bc-2\b\g)\\-c(j^W_B+j^{K'}_B+j^{co}_B)+\frac12
cb\g+x\pa\g \end{array}\right](z)\nonumber\\
&[\tilde{B}^W+B^{K'}+B^{co}+2\b\pa c+c\pa\b+\m b](w )\nonumber\\
&\sim (cb\g-\b\g\g+x\pa\g)(z)(2\b\pa c+c\pa\b)(w)\nonumber\\
&-c(z)[j^W_B(z)\tilde{B}^W(w)+j_B^{K'}(z)B^{K'}(w)+j_B^{co}(z)B^{co}(w)]+\dots\nonumber \\
&\sim \frac{3c(z)}{(z-w)^3}-\frac{2xc(w)}{(z-w)^3}-\frac{2x\pa
c(w)}{(z-w)^2}-\frac{c(z)}{(z-w)^3}[-22-3-6]+\dots
\end{align}
The triple poles cancel for x=17, confirming again the result for $x$.
Then also the double poles cancel, while from the simple poles we
find that $\gp$ contains a term $-17\pa^2 c$. This yields again
$y=-17$.

\section*{Acknowledgements}

It is a pleasure to thank L. Castellani, A. Lerda and 
the Department of Science of Piemonte Orientale 
University at Alessandria for financial support 
during our stay, E. Verlinde for drawing our attention 
to reference \cite{oldN=4} and S. Guttenberg and M. Kreuzer 
for correspondence.  
P.A.G. thanks the organizers of the Second Simons Workshop in 
Mathematics and Physics at Stony Brook. 
This works was supported by NSF grant PHY-0098527.


\end{document}